\documentstyle[12pt]{article}
\newcommand{\e}[1]{\label{eq:#1}}
\newcommand{\ee}[1]{(\ref{eq:#1})}

\newcommand{\eq}{\begin{equation}}
\newcommand{\eqe}{\end{equation}}
\newcommand{\eqa}{\begin{eqnarray}}
\newcommand{\eqae}{\end{eqnarray}}

\newcommand{\del}{\partial}
\newcommand{\zb}{{\bar{z}}}

\newcommand{\ket}[1]{\mbox{$| #1 \rangle$}}

\newcommand{\av}[1]{{\tt <} #1 {\tt >}}

\begin{document}
\hfill{UTTG-16-92}

\hfill{June, 1992}

\vspace{24pt}
\begin{center}
{\bf Strings and QCD?}

\vspace{24pt}
Joseph Polchinski

\vspace{12pt}
Theory Group\\ Department of Physics \\ University of Texas
\\ Austin, Texas 78712\\ bitnet:joe@utaphy

\vspace{6pt}

\end{center}
\baselineskip=15pt

\begin{minipage}{4.8in}
{\bf Abstract:}
Is large-$N$ QCD equivalent to a string theory?  Maybe, maybe not.
I review various attempts to answer the question. \\

Presented at the Symposium on Black Holes, Wormholes, Membranes, and
Superstrings, H.A.R.C., Houston, January 1992.
\end{minipage}

\vspace{24pt}

It is an old and seductive idea that large-$N$ QCD might be exactly
reformulated as a string theory[1].
It is probably too much to hope that the large-$N$ theory can be solved
directly,\footnote
{Shortly after writing these words, I learned of the recent work by Migdal
and Kazakov, reducing gauge theory to a scalar
matrix theory which can be solved by matrix model methods.  This is a very
clever idea, though my initial
reaction is skepticism that the model that has been solved is really
QCD: the non-Abelian information that one expects in the master field
is not there.   But perhaps this objection is naive; certainly this deserves a
closer look, and might render my talk irrelevant!}
but the topological structure of the perturbation theory and
the success of Regge phenomenology make it plausible that one could first
recast it as a string theory, and then apply string methods to determine
the spectrum and amplitudes.  Over the years a fair amount of work
has gone into this, but it is not clear that we have learned a single
thing about QCD.  Fortunately the effort has not been wasted, because
it has led to many discoveries
which have been important to the string theory of the fundamental
interactions.

Since 1984, several thousand physicist-years have gone into the
development of string theory.  It therefore seems timely to reexamine the
question, ``What, if anything, is the relation between strings and QCD?''
In recent months I have thought about this from a number of points of view.
The result is something like the parable of the blind men and the elephant:
there is very little overlap between the different perspectives.  In the next
six sections I will describe the elephant as seen by six of these blind men.

\section{Regge Phenomenology}

One often encounters the statement ``Regge phenomenology works better
than it ought to.''  As one example, the trajectories remain fairly
straight down to small angular momentum.  As a more detailed example,
applying dual resonance theory and PCAC to the process
$\pi + A \to A^*$(resonance),
leads to the mass relations[2]
\eq
m^2(A^*) - m^2(A) = \frac{1}{2\alpha'} \times \mbox{odd integer}. \e{1}
\eqe
A number of relations of this form hold to within a few per cent,
for example with $A^* = \rho$ and $A = \pi$.   This is striking
evidence that the string picture is good even down to the pion.

Llewellyn has
recently updated the data and tried to find an explanation of the result~\ee{1}
within modern conformal field theory[3].  He makes a nice try: if one has an
$SU(2)_R \times SU(2)_L$ current algebra on the string world-sheet, boundaries
will break this down to $SU(2)_D$.  From the spacetime point of view, this is
spontaneous breaking, and gives rise to a massless Goldstone state.  Now,
by the usual rules, the $SU(2)_L \times SU(2)_R$ symmetry is a local
symmetry in spacetime, with gauge bosons from the closed string sector,
and the Goldstone boson is eaten.  So we must imagine that there is some way to
change the rules and get rid of these gauge bosons, in which case there
is now a Goldstone pion.  Llewellyn argues that the mass
relations~\ee{1} then follow from an operator product locality argument.
It seems very reasonable to play with the idea of keeping some of the usual
rules while ignoring others, but it is hard to reconcile this argument with
the picture one gets from large-$N$ QCD.  The $SU(2) \times SU(2)$ chiral
symmetry is carried by the quarks but not the gluons.  In the usual string
picture, based on the large-$N$ approximation, the quarks live only at the
endpoints of the string, so that the $SU(2) \times SU(2)$ should be an endpoint
(Chan-Paton) symmetry, not an interior (Kac-Moody) symmetry: these are quite
different, and the former does not seem to work for the present purpose.

It is interesting to consider the splitting~\ee{1} further.  The $\rho$
and $\pi$, differ only in the
relative orientation of the quark and antiquark spins.  These live at
opposite ends of the string and communicate only
through their couplings to the string between.  Evidently there is
something in this coupling that gives rise to universal splittings.
How do the quark spins couple to the string
variables?  Unfortunately, no one seems to have found a string model in
which the endpoints carry spin quantum numbers, and I have not found
anything simple.  This would seem to be a prerequisite for string Regge
phenomenology, since the data is all for mesons and baryons, not glueballs.

\section{Loop Equations}

Loop expectation values would seem to be the natural observables both in
gauge theory and string theory, and an obvious starting point in trying
to relate the two[4].  In gauge theory the Wilson loop is the basic geometric
observable, measuring the curvature of the field.  In string theory, the sum
over world-sheets with fixed boundary loops defines a natural correlation
function.  Unfortunately, this plausible approach quickly becomes mired in
technicalities.  In gauge theory, the field equations for Wilson loops are
notoriously ill defined.  In string theory, the path integral with fixed
boundary loops turns out to be difficult to evaluate, and to have a
complicated and poorly understood analytic structure.  Other string
observables---the vertex operators and the BRST-invariant string fields---are
far simpler, and their world-sheet and spacetime analytic and symmetry
properties much clearer.  Perhaps with some insight these difficulties can be
resolved.  Even in matrix models, where in principle everything is solvable,
the dictionary between loop observables and string variables has proven quite
complicated[5].  This is perhaps a place where matrix models will eventually
teach us something useful to higher-dimensional physics.

The loop equations for gauge theory have
sometimes been argued to be equivalent to those for one or another
specific string theory, but the above complications make it difficult
to evaluate these claims.
Some of these ideas can be ruled out by study of the high energy behavior
of the partition function, to be discussed below.

\section{Lattice Strong Coupling Expansion}

Wilson's strong coupling expansion for lattice gauge theory strongly
resembles a string theory[6].  Terms involving non-intersecting
sets of plaquettes look {\it exactly} like string theory: a sum over
surfaces weighted by the exponential of the area.  This led to some
optimistic conjectures, but Weingarten pointed out that the weight for
self-intersecting surfaces is more complicated: it implies a contact
interaction[7].  Contact interactions, such as self-avoidance
\eq
X^\mu(\sigma) \neq X^\mu(\sigma') \quad\mbox{for all $\sigma$, $\sigma'$}
\eqe
are non-local on the world-sheet and so represent a great complication of
the theory.  Subsequently it was shown that, as for the two- and
three-dimensional Ising models, the theory could be written as a sum over
surfaces with a local weight but with additional world-sheet variables[8].

This result is very suggestive.  Essentially it is an existence proof for
a string theory of QCD.  Of course, the strong coupling expansion is far
from the continuum limit, but because the strong coupling expansion has a
finite radius of convergence, it seems reasonable that an equality which holds
to all orders in the strong coupling expansion will hold in some form when
continued to the continuum limit.  The new world-sheet
variables are infinite in number, and the modified theory rather
complicated, so it is hard to guess the form of the continuum
string theory.  At least some of the new variables seem to be associated
with the weight given to special
points, like saddles, on the string world-sheet.  (Also, to
answer a question raised by A. Strominger, the $1/N$ expansion is more
complicated than might be expected.  The new variables are classical at
$N=\infty$ but fluctuate at finite $N$, so there is a $1/N$ expansion
on the world-sheet in addition to the topological expansion.)

\section{Two-Dimensional Gauge Theory}

Two-dimensional gauge theory is an interesting arena for string ideas.
Consider a Wilson loop $W_R(L)$, where $L$ is a non-self-intersecting loop
and $R$ denotes the representation.  The expectation value is readily
found to be
\eq
< W_R(L) > \ =\ e^{-C_2(R) A_L/2g^2}
\eqe
from the energy of the electric flux, with $C_2(R)$ being the Casimir and
$A_L$ the area of $L$.  This looks rather stringy.  For two such loops,
separated from one another, the expectation value simply factorizes (for
all $N$),
\eq
< W_R(L_1) W_R(L_2) > \ =\ e^{ -C_2(R) A_1/2g^2 - C_2(R) A_2/2g^2}.
\eqe
This factorization occurs because there are no propagating gluons.  However,
from the string point of view it is a bit strange, because the cylindrical
world-sheet bounded by $L_1$ and $L_2$ might be expected to make a non-zero
contribution and therefore give a correlation.  Noting that the cylinder must
be squashed to fit into two dimensions, we might imagine that the string is
self-avoiding, thus forbidding this world-sheet.  This is not quite right,
though, because there are examples (overlapping loops, and self-intersecting
loops), where this rule would not work.  Also, we have noted in the previous
section that self-avoidance is non-local on the world-sheet, and a strictly
local interaction is both desirable and possible.  The squashed cylinder
can be forbidden instead by the local rule that {\it folds} on the world-sheet
are not permitted.
Kazakov and Kostov made a detailed study of expectation values of Wilson
loops in two dimensions[9], and Kostov showed that on the lattice the results
can be obtained from the strong coupling sum over surfaces theory described in
the previous section[10] (except that the proof of the no-folds rule is
not complete). Recently, gauge theories on topologically nontrivial
two-dimensional spacetimes have also been considered[11].  It would be an
interesting exercise to see how the lattice strong-coupling expansion
reproduces these.

\section{Long Strings at Low Energy}

In this section and the next I wish to consider a gedanken
experiment.\footnote
{For further details and references, see ref.~[12] for the present section
and~[13] for the next.}
The experiment is rather tame by the standards of this meeting:
no black holes will be involved.  I simply wish to grab hold of a QCD string
and
stretch it out (imagining that quarks are absent or heavy, so that the string
cannot break) and then to shake it and squeeze it,
and to compare the results with the same
experiments on a fundamental string.  A basic quesion is the number of degrees
of freedom in each case.  In the present section I will consider low
frequencies
(compared to the scale set by the string tension) and in the next section
high frequencies.

Obviously, low frequencies are less revealing than high, but there is an old
and
interesting paradox to deal with.  A long QCD (or Nielsen-Oleson) string should
be described to first approximation by the Nambu-Goto action, and its
excitations by some quantization of this action.  The QCD and Nielsen-Oleson
(NO) strings exist within consistent quantum
field theories---how does this fit with
the problem of quantizing the Nambu-Goto action outside the critical
dimension?   First we must sharpen the paradox, and see that the problem of the
critical dimension arises even in the long-string limit where the Nambu-Goto
description should be valid.  The basic problem is that none of the standard
quantizations gives the correct spectrum for a QCD or NO string in this limit.
The old covariant (Virasoro) quantization starts with $4$ oscillators, where
I have specialized to $D=4$ where the QCD and NO strings live.
Classically, world-sheet coordinate invariance
reduces this to $2$, but below the critical dimension an anomaly in this
invariance reduces the number of null states and leaves roughly $3$ sets of
oscillators.  In contrast, the light-cone quantization gives only the $2$
transverse oscillators, but the spectrum is not Lorentz invariant outside the
critical dimension.  Polyakov introduced two new ideas[14]: an independent
world-sheet metric field, and a careful treatment of the path integral
measure.  These nearly offset one another: after gauge-fixing there is one
additional world-sheet scalar, the Liouville field, for $5$ in all,
but now the coordinate algebra is non-anomalous reducing the number to
$3$.  In fact, except for the subtlety of the Liouville zero mode, the
old covariant and Polyakov quantizations are the same[15].

How do these results compare with what is expected for the QCD and NO
strings?  From the point of view of the string world-sheet,
the oscillators correspond to massless scalars.  On general field-theoretic
principle, massless scalars are natural only if they are Goldstone bosons.
The $2$ transverse oscillations are indeed of this type, but there is no
place for the $3$'rd to come from.  For the
NO string, one can see explicitly from the semiclassical expansion that there
are only $2$ degrees of freedom.  For the QCD string one cannot, but it seems
overwhelmingly likely on grounds of naturalness.  Of the standard
quantizations, only the light-cone quantization gives the correct number of
degrees of freedom, but it is noncovariant while the QCD and NO strings live
within covariant theories.\footnote{The spontaneous breaking of Lorentz
invariance by a long string should not be confused with the explicit breaking
in the light-cone quantization.  See ref.~[16] for further
discussion.}

For the NO string, one can in principle find the correct quantization,
starting from the quantization of the underlying gauge theory, by careful
treatment of the collective coordinate quantization.  This has not been done,
but it is not too hard to guess what the result should be.  The old covariant
and light-cone quantizations have always seemed suspect from a path integral
point of view, because one carries out classical manipulations without regard
for measure factors before quantizing.  One should be
careful about the measure throughout, as did Polyakov.  On the other hand, his
second innovation, the independent metric, will not come out of the collective
coordinate method: lengths should be set by the spacetime metric, through the
induced world-sheet metric.  The guess is therefore that one will obtain
Polyakov's determinants but in terms of the induced rather than Liouville
metric---let me call this the effective string.  One can then check that
this gives the expected spectrum: the coordinate invariance is non-anomalous
(matter central charge = 26) but  there are only $4$ fields and so $2$
physical sets of oscillators. One can also confirm this result in another way,
writing the most general long-string CFT with $4$ $X^\mu$ fields and requiring
$c = 26$.

Notice that this is certainly not the same as the Polyakov/Liouville
quantization, as in the latter the Liouville field is decoupled from the
embedding.  Rather it has in effect a Lagrange multiplier setting the induced
and intrinsic metrics equal.

What do we learn?  First, that the Liouville theory for $c > 1$, while an
interesting problem in field theory, is not relevant to physical (QCD
or NO) strings.  This has also been argued by others, such as Banks, from
other points of view.  Second, it is interesting that the result,
a covariant Poincar\'e invariant $c=26$ CFT constructed from 4 $X^\mu$
fields, can be proved not to exist!  To see this, consider the world-sheet
current of spacetime translation invariance, $j^\mu_a$.  Conformal invariance
determines
\eq
\av{ j^\mu_z(z) j^\mu_z(-z) } \propto z^{-2}, \e{jj}
\eqe
whence
\eqa
0 &=& \av{\del_\zb j^\mu_z(z) \del_\zb j^\mu_z(-z)} \nonumber\\[2pt]
&=& \| \del_\zb j^\mu_z(z) \ket{0} \|^2.
\eqae
But a local operator which annihates the vacuum is zero, which means
that the current is analytic[17].  This is sufficient to show that the $X^\mu$
CFT is free; together with Lorentz invariance, this fixed the energy-momentum
tensor to the standard form $-\del_z X^\mu \del_\zb X_\mu/2$
and so has $c=4$.  In the present case, it is the first
step~\ee{jj} which goes wrong, because conformal invariance is spontaneously
broken in the long-string background by the expectation value of $\del_a
X^\mu$.\footnote{Spontaneous breaking of conformal invariance is the
identifying feature of non-critical strings.}  Thus, the translation currents
are not analytic, and the $X^\mu$ are not free.

The effective string action
involves the inverse of $\del_a X^\mu \del_b X_\mu$ and so only makes sense in
the long-string background.  It breaks down for short (low-lying) string
states.  Also, because it is nonrenormalizable, it breaks down for long
strings shaken at high frequency.  So it cannot be a complete theory, but only
a clue.  Incidentally, it is hard to violate the above theorem: I am aware of
two renormalizable string theories whose low energy limit is the effective
string, and both are pathological: the rigid string[18], which as I will
discuss
below is non-unitary, and a long string in a certain class of sigma model
backgrounds, considered by Natsuume[19], which necessarily
has world-sheet tachyons.

\section{Long Strings at High Energy}

As we shake the string at higher and higher frequencies, we would not be
surprised to excite new world-sheet degrees of freedom beyond the $2$
transverse
oscillators.  One thing we have learned in recent years is that the string
partition function is very revealing, and often simpler to evaluate than the
interactions.  Let us try to count the number of degrees of freedom of the
string via the partition function.  In QCD at very high temperatures, we might
expect that this would be possible due to asymptotic freedom.  The problem is
that there is a transition to a deconfining phase at high temperature.
This transition is very similar to the Hagedorn transition in string theory:
a winding state in the Euclidean temperature direction becomes
tachyonic and acquires an expectation value.\footnote{We are concerned here
with analytic continuation of the canonical ensemble; the actual physics at the
string Hagedorn temperature may be more complicated.}  Now, in string
perturbation theory we routinely ignore tachyons, expanding around the
maximum in the potential.  It therefore
seems plausible that this should also be
possible in QCD, corresponding to analytic continuation of the confined-phase
partition function to high temperature.  Indeed, this turns out to be
extremely easy, given existing results on the high-temperature effective
potential.  I will not repeat here the details which appear in ref.~[13].
For the string free energy per unit length, one finds in QCD
\eq
\mu^2(\beta)/L^2 \sim -\frac{2g^2(\beta) N}{\pi^2 \beta^4}, \e{qcd}
\eqe
to be compared with the Nambu-Goto spectrum with $n$ sets of oscillators,
\eq
\mu^2(\beta)/L^2 \sim -\frac{n}{6 \alpha' \beta^2}.
\eqe
These do not agree unless more and more fields are excited as the
temperature increases, $n_{\rm eff}(\beta) \propto \beta^{-2}$.

The are several possible interpretations.  The simplest is that the QCD string
is simply a fat object.  The number of internal shape excitations would indeed
be expected to scale as energy-squared (like the spectrum of a field in a
two-dimensional cavity).  This is really the essential issue: large-$N$
perturbation theory is planar in index space, but is there any sense in which
it is two-dimensional (= thin string) in its spacetime structure?  If not, the
string picture is unlikely to be useful outside of the long-string and Regge
limits.
Fortunately, there are indications that the QCD string is not as complicated
as it might be.  The first is the lattice strong-coupling expansion, which is
at least formally a representation of QCD in terms of sums over thin surfaces
with additional fields.  It is tempting to identify the extra fields of the
strong coupling expansion with those found in the partition function, but I
do not have a direct connection between these.\footnote{Note, though, that
the partition function rules out various proposals that large-$N$ QCD is
equivalent to a Nambu-Goto string with a {\it finite} number of additional
degrees of freedom.}  The prospect of an infinite number of fields might in any
case make one worry that a string description will be too complicated to
be useful; however, if one is interested in low-lying hadrons it may be that
all but a few of the fields decouple.

Another indication that the large-$N$ QCD string is thin comes from further
application of the high-temperature continuation.  If the string is squeezed
by imposing transverse periodic boundary conditions, its spectrum and
correlations do not change (this follows from the planarity of the large-$N$
approximation).  The simplest interpretation is that it does not feel the
squeezing because it is a thin object.  Incidentally, this is equivalent to the
Eguchi-Kawai reduction: by compactifying all dimensions spacetime is reduced
to a point without changing the correlations[19]!  (The high-temperature
continuation is equivalent to what is referred to as `quenching.')  This seems
remarkable, but the mechanics of it is quite simple.  Focus on the case of
one compact dimension ($0 < \tau < \beta$).  In a gauge in which the $\tau$
component of the vector potential is diagonal and $\tau$-independent,
\eq
A^\tau(\tau) = {\rm diag}(A^\tau_i),
\eqe
the confined phase is one in which the Polyakov loops average to zero, so the
phases $e^{ i\beta A^\tau_i }$ are distributed uniformly around the circle.
Now, if we take $\beta \to 0$ and dimensionally reduce,
the covariant derivative is
\eq
D_\tau = \del_\tau + i A^\tau \to A^\tau.
\eqe
In particular, an $ij$ gluon has covariant momentum $A^\tau_i - A^\tau_j$.
Because the $A^\tau_i$ are uniformly distributed, the index sum $i$ around
each loop effectively produces a momentum integral!  By taking $N$
sufficiently large, that is, the momentum integral can be hidden in
the color sum.
Once this is understood, we see that the reduction is bit of a swindle,
and indeed it has not as yet led to useful progress.

Unfortunately, what we have calculated for QCD cannot be directly interpreted
in
terms of a number of degrees of freedom because it has been continued to an
unphysical region.  Above, we used the Nambu-Goto spectrum for comparison, but
there are modifications of string theory which change the high-energy
behavior.  (This point was made to me independently by Zhu Yang and Mike
Green).
One is the `rigid string,' where the world-sheet ultraviolet behavior is
dominated by an
extrinsic curvature-squared term.  Yang and I have calculated the continued
partition function for the rigid string, and find that it gives the same power
of temperature as QCD, eq.~\ee{qcd}, though the sign is wrong[21].  However,
the
rigid string has a serious and basic problem which precludes it from being of
any relevance to QCD: the curvature term gives the $X^\mu$ a fourth derivative
action, so the spectrum is not unitary.  As far as we can determine, this
problem has been almost totally ignored in the literature (except of
ref.~[22]), but if the curvature term dominates at any scale, then there must
be negative norm (or negative energy) modes at that scale.  So this theory
cannot apply to strings in Minkowski space.  Of course, it could (and evidently
does) apply to the statistical mechanics of surfaces, because there is no
Minkowski continuation here.

Another short-distance modification is the introduction of boundary loops
with Dirichlet conditions ($X^\mu$ constant on a given boundary, the value of
the constant being integrated over).  These are pointlike in spacetime but
extended in terms of the world-sheet conformal structure, and have a
substantial effect on the behavior of amplitudes.  They were proposed in order
to introduce partonic structure into string theory, and Mike Green in
particular has pursued this over the years[23].  Recently he has shown that
these appear to give the correct behavior for the continued partition
function. I find this idea very intriguing: it has some of the correct
properties, is nontrivial enough that it could be correct, but at the same time
is simple enough that it might be useful.  In any case, consistent
modifications of string theory are few and far between, so it is of interest to
see whether this is consistent, and what the physics of it is.

\section{Conclusions}

The initial goal was to see whether recent years' progress in string theory
enables us to evaluate the conjecture that large-$N$ QCD is a string theory.
Indeed, we have learned a few things, though the results are not especially
favorable.  The study of the long string low energy limit revealed that the
embedding coordinates $X^\mu$ cannot be free fields.  By contrast, for the
reason discussed in section~5 they {\it are} free in all Poincar\'e
invariant four-dimensional
critical string theories.  This is a great complication: the simplicity of
dual models comes largely because they are built from free
oscillators.\footnote{McGuigan and Thorn have recently made the clever
observation that Regge trajectories at large negative $t$ can be evaluated in
QCD perturbation theory[24], somewhat similar in spirit to the idea in
section~6. They do not remain linear, but approach constants.  This lack of
linearity also shows that the $X^\mu$ are not free.}
The study of the long string high temperature limit highlights the differences
between string theory and QCD at short distance.  Of course this difference is
well-known in the interactions (soft versus partonic), but this shows that
it can also be seen in the partition function, which has been
a very useful object in string theory.

Perhaps these differences are a sign that we are at a dead end.  However, I
think that there is still tantalizing evidence that there is something
more to be said.  The best approach I see is to try to guess the continuum
limit of the strongly coupled lattice theory, perhaps aided by the continuum
high temperature continuation/Eguchi-Kawai reduction.  Also, I would like to
better understand the physics of Dirichlet boundaries.\\

{\bf Acknowledgements:}
I would like to thank D. Minic, M. Natsuume, E. Smith, A. Strominger, and
Z. Yang for discussion and collaboration on some of these points of view.
This work was supported in part by the Robert A. Welch Foundation, NSF Grant
PHY 9009850, and the Texas Advanced Research Foundation.

\vfill

\pagebreak

\centerline{\bf References}

\begin{itemize}

\item[1.] G. 't Hooft, Nucl. Phys. {\bf B72}, 461 (1974).

\item[2.] C. Lovelace, Phys. Lett. {\bf B34}, 500 (1971);
M. Ademollo, G. Veneziano, and S. Weinberg, Phys. Rev. Lett. {\bf 22},
83 (1969).

\item[3.] D. C. Lewellyn, ``Effective String Amplitudes for Hadronic Physics,''
ITP preprint NSF-ITP-91-105 (1991).

\item[4.] For a review see A. A. Migdal, Phys. Rep. {\bf 102}, 199 (1983).

\item[5.] See, for example, G. Moore, N. Seiberg, and M. Staudacher,
Nucl. Phys. {\bf B362}, 665 (1991).

\item[6.] K.G. Wilson, Phys. Rev. {\bf D10}, 2445 (1974).

\item[7.] D. Weingarten, Phys. Lett. {\bf B90}, 285 (1980).

\item[8.] V. A. Kazakov, Phys. Lett. {\bf B128}, 316 (1983);
V. I. Kostov, Phys. Lett. {\bf B138}, 191 (1984);
K. H. O'Brien and J.-B. Zuber, Nucl. Phys. {\bf B253}, 621 (1985).

\item[9.] V. Kazakov and I. Kostov, Nucl. Phys. {\bf B220}, 167 (1983).

\item[10.] I. K. Kostov, Nucl. Phys. {\bf B265}, 223 (1986).

\item[11.] M. Blau and G. Thompson, ``Quantum Yang-Mills Theory on Arbitrary
Surfaces,'' preprint NIKHEF-H/91-09, MZ-TH/91-17;
E. Witten, ``On Quantum Gauge Theories in Two Dimensions,'' Comm. Math. Phys.
{\bf 141}, 153 (1991);
A. Strominger, unpublished.

\item[12.] J. Polchinski and A. Strominger, Phys. Rev. Lett. {\bf 67}, 1681
(1991).

\item[13.] J. Polchinski, Phys. Rev. Lett. {\bf 68}, 1267 (1992).

\item[14.] A. M. Polyakov, Phys. Lett. {\bf B40}, 235 (1972).

\item[15.] R. Marnelius, Phys. Lett. {\bf B172}, 337 (1986).

\item[16.] J. D. Cohn and V. Periwal, ``Lorentz Invariance of
Effective Strings,'' IAS/Fermilab preprint IASSNS-HEP-92-18,
Fermilab-92/126-T (1992).

\item[17.] R. Dashen and Y. Frishman, Phys. Rev. {\bf D11}, 2781 (1975);
I. Affleck, Phys. Rev. Lett. {\bf 55}, 1355 (1985).

\item[18.] A. M. Polyakov, Nucl. Phys. {\bf B268}, 406 (1986);
H. Kleinert, Phys. Lett. {\bf B174}, 335 (1986).

\item[19.] M. Natsuume, ``Nonlinear Sigma Model for String Solitons,''
Texas preprint UTTG-10-92 (1992).

\item[20.] T. Eguchi and H. Kawai, Phys. Rev. Lett. {\bf 48}, 1063 (1982);
D. J. Gross and Y. Kitazawa, Nucl. Phys. {\bf B206}, 440 (1982);
A. A. Migdal, Phys. Lett. {\bf B116}, 425 (1982);
S. Das and S. Wadia, Phys. Lett. 117B (1982) 228;
see also ref.~[4].

\item[21.] J. Polchinski and Z. Yang, ``High-Temperature Partition Function of
the Rigid String,'' Texas/Rochester preprint UTTG-08-92, UR-1254, ER-40685-706
(1992).

\item[22.] E. Braaten and C. K. Zachos, Phys. Rev. {\bf D34}, 1512 (1987).

\item[23.] M. B. Green, ``Temperature Dependence of String Theory in the
Presence of World-Sheet Boundaries,'' Queen Mary College preprint QMW-91-24
(1991).

\item[24.] M. McGuigan and C. T. Thorn, ``Quark-Antiquark Regge Trajectories
in Large-$N_c$ QCD,'' Florida preprint UFIFT-92-12 (1992).

\end{itemize}

\end{document}